\documentclass[twocolumn,showpacs,preprintnumbers,amsmath,amssymb]{revtex4}

\usepackage{graphicx}
\usepackage{dcolumn}
\usepackage{bm}

\begin{document}

\title{The process $\mu \rightarrow \nu _{e}e\overline{\nu }_{\mu }\;$in the 2HDM
with flavor changing neutral currents.}

\author{Rodolfo A. Diaz, R. Martinez and Nicanor Poveda }
\affiliation{Universidad Nacional de Colombia, Departamento de Fisica.\\
Bogota, Colombia.}         

\date{\today}

\begin{abstract}
We consider the process $\mu \rightarrow \nu _{e}e\overline{\nu }_{\mu }\;$%
in the framework of a two Higgs doublet model with flavor changing neutral
currents (FCNC). Since FCNC generates in turn flavor changing charged
currents in the lepton sector, this process appears at tree level mediated
by a charged Higgs boson exchange. From the experimental upper limit for
this decay, we obtain the bound $\left| \xi _{\mu e}/m_{H^{\pm }}\right|
\leq 3.8\times 10^{-3}$ where$\;\xi _{\mu e}\;$refers to the mixing between
the first and second lepton generations, and $m_{H^{\pm }}\;$denotes the
mass of the charged Higgs boson. This bound is independent on the other free
parameters of the model. In particular, for $m_{H^{\pm }}\simeq 100$GeV\ we
get $\left| \xi _{e\mu }\right| $ $\lesssim 0.38$
\end{abstract}


\maketitle

Flavor Changing Neutral Currents (FCNC) are processes highly suppressed by
some underlying principle still unknown. However, the observation of new
phenomena like oscillation of neutrinos \cite{Beshtoev} coming from the sun
and the atmosphere, seem to indicate the existence of such rare couplings.
Neutrino oscillations can be explained by introducing mass terms for these
particles, in whose case the mass eigenstates are different from the
interaction eigenstates. It should be emphasized that the oscillation
phenomenon implies that the lepton family number is violated, and such fact
leads us in turn to consider the existence of physics beyond the standard
model (SM), because in SM neutrinos are predicted to be massless and lepton
flavor violating mechanisms are basically absent. These considerations
motivate the study of scenarios with Lepton Flavor Violation (LFV) and
Family Lepton Flavor Violation (FLFV).

The original motivation for the introduction of neutrino oscillations comes
from the first experiment designed to measure the flux of solar neutrinos 
\cite{Davis}, such measurement was several times smaller than the value
expected from the standard solar model, so ref \cite{Pontecorvo} suggested
the neutrino oscillation mechanism as a possible explanation of the neutrino
deficit problem. In addition, models of neutrino oscillations in matter \cite
{Mikheyev} arose to solve the neutrino deficit confirmed by
SuperKamiokande \cite{Hirata}. Since then, further evidence of solar
neutrino oscillations has been found by SuperKamiokande and SNO \cite{Kameda}%
. Besides, this phenomenon can be inferred from experiments with atmospheric
neutrinos as well \cite{Fukuda}.

On the other hand, since neutrino oscillations imply FLFV in the neutral
lepton sector, it is generally expected to find out FLFV processes involving
the charged lepton sector as well, searches for these processes have
provided some upper limits for several decays involving these exotic
mechanisms, some examples are $\mu -e\;$conversion in nuclei \cite{doh}, $%
\mu \rightarrow eee\;$\cite{bell}$,\;\mu \rightarrow e\gamma \;$\cite{bolton}%
$,\;$and $\mu ^{-}\rightarrow \nu _{e}e^{-}\overline{\nu }_{\mu }$\cite
{Okada}. In addition, the search for FLFV can also be made by analyzing
semileptonic processes, upper bounds from FLFV meson decays have been
estimated, some examples are$\;K_{L}^{0}\rightarrow \mu ^{+}e^{-}$ \cite
{arisaka}, and $K_{L}^{0}\left( K^{+}\right) \rightarrow \pi ^{0}\left( \pi
^{+}\right) \mu ^{+}e^{-}$ \cite{plb}. Moreover, the phenomenon of LFV has
been widely studied from the theoretical point of view in different
scenarios such as Two Higgs Doublet Models, Supersymmetry, and Grand
Unification \cite{nosno}.

With these motivations in mind, we shall study the FLFV decay $\mu
^{-}\rightarrow \nu _{e}e^{-}\overline{\nu }_{\mu }$ which provides
information about the mixing between the first and second lepton family. We
examine this process in the framework of one of the simplest extension of
the standard model that generates FLFV at tree level, the so called Two
Higgs Doublet Model type III (2HDM (III)). This model consists of adding a
second doublet to the SM with the same quantum numbers of the first, and
considering all type of Yukawa couplings. Some bounds on lepton flavor
violating couplings involving the mixings $\mu -\tau ,\;e-\tau ,\;$and $\tau
-\tau \;$have been considered recently \cite{nos} by taking into account the 
$g-2$ muon factor and some of the leptonic decays cited above. Ref \cite{nos}
considers processes with Flavor Changing Neutral Currents (FCNC) to study
LFV. Instead, the process $\mu ^{-}\rightarrow \nu _{e}e^{-}\overline{\nu }%
_{\mu }\;$to be study in this brief report, involves Flavor Changing Charged
Currents (FCCC) and the mixing between the first and second lepton family.

The Lepton sector in the Yukawa Lagrangian type III reads 
\begin{eqnarray}
-\pounds _{Y} &=&\overline{E}\left[ \frac{g}{2M_{W}}M_{E}^{diag}\right]
E\left( \cos \alpha H^{0}-\sin \alpha h^{0}\right) \nonumber \\
&+&\frac{1}{\sqrt{2}}\overline{E}\xi ^{E}E\left( \sin \alpha H^{0}+\cos
\alpha h^{0}\right) \nonumber \\
&+&\overline{\vartheta }\xi ^{E}P_{R}EH^{+} +\frac{i}{\sqrt{2}}%
\overline{E}\xi ^{E}\gamma _{5}EA^{0}  + h.c.
\label{lag}
\end{eqnarray}
where $H^{0},h^{0}$\ denote$\;$the heaviest and lightest scalar Higgs bosons
respectively. $A^{0}\;$is a pseudoscalar Higgs boson, and $E,\;\left(
\vartheta \right) \;$refers to the three charged, (neutral) leptons i.e. $%
E\equiv \left( e,\mu ,\tau \right) ^{T},\;\vartheta \equiv \left( \nu
_{e},\nu _{\mu },\nu _{\tau }\right) ^{T}$. The matrices $M_{E},\;\xi _{E}\;$%
describe the lepton masses and the flavor changing vertices respectively.
Finally, $\alpha \;$is the mixing angle in the scalar Higgs bosons sector.
We shall deal with a parametrization in which only one of the doublets
acquires a vacuum expectation value.

From Lagrangian (\ref{lag}) we see that the matrix elements $\xi _{ij}^{E}\;$%
that generates FCNC automatically generates FCCC which are strongly suppress
in the leptonic sector. Consequently, by constraining flavor changing
charged currents we are indirectly constraining flavor changing neutral
currents as well. In the case of the 2HDM (III), interactions involving FCCC
at tree level only contains the contribution from the charged Higgs boson,
reducing the free parameters to manage.

In this note, we shall extract a bound for the quotient $\xi _{e\mu
}/m_{H^{+}}\;$based on the constraints on the three body decay $\mu
^{-}\rightarrow \nu _{e}e^{-}\overline{\nu }_{\mu }\;$mediated by a charged
Higgs, this decay produces FLFV. The corresponding decay width is given by

\begin{equation}
\Gamma \left( \mu \rightarrow \nu _{e}e^{-}\overline{\nu }_{\mu }\right) =%
\frac{m_{\mu }^{5}}{24\,576\pi ^{3}}\left( \frac{\xi _{12}}{m_{H}}\right)
^{4}
\end{equation}
and taking the current upper bound for this decay \cite{data particle} 
\begin{equation}
\Gamma \left( \mu ^{-}\rightarrow \nu _{e}e^{-}\overline{\nu }_{\mu }\right)
\leq 3.\,\allowbreak 6\times 10^{-21}GeV,
\end{equation}
the following constraint is gotten 
\begin{equation}
\left| \frac{\xi _{e\mu }}{m_{H^{+}}}\right| \leq 3.8\times 10^{-3}GeV^{-1}.
\end{equation}
Despite this constraint is not so strong, it is interesting since it does
not depend on the other free parameters of the model, because the
calculation does not involve neutral Higgs bosons nor mixing angles. This is
a good motivation to improve the experimental upper limit for processes
involving FCCC in the leptonic sector.

In conclusion, we can constrain flavor changing neutral currents in the 2HDM
(III) also by examining flavor changing charged currents. The study of the
latter at tree level in the 2HDM (III) depend on a less number of free
parameters since only one scalar particle is exchanged, and the Higgs boson
involved does not couple through a mixing angle. As a manner of example, for 
$m_{H^{\pm }}\simeq 100$ GeV\ we obtain $\left| \xi _{e\mu }\right| $ $%
\lesssim 0.38$\ without any further assumption.

This work was supported by COLCIENCIAS and Fundaci\'on Banco de la Rep\'ublica.

\end{document}